\documentclass{aa}

\usepackage{graphicx}
\usepackage{txfonts}
\usepackage{lipsum}
\usepackage{subcaption}         
\usepackage{lscape}             
\usepackage{placeins}           
\usepackage{amsmath}            
\usepackage{xspace}             
\usepackage{xcolor}             

\newcommand{\software}[1]{\texttt{#1}}
\newcommand{\unit}[1]{\ensuremath{\mathrm{\,#1}}\xspace}
\newcommand{\kms}{\unit{km\,s^{-1}}}

\newcommand{\kpc}{\unit{kpc}\xspace}
\newcommand{\gaia}{{\it Gaia}\xspace}

\newcommand{\teff}{\ensuremath{T_{\mathrm{eff}}}}
\newcommand{\logg}{\ensuremath{\log(\mathrm{g})}}
\newcommand{\feh}{\unit{[Fe/H]}}
\newcommand{\sag}{Sgr~A$^\ast$\xspace}
\newcommand{\candidate}{DESI-312\xspace} 

\newcommand{\SSSSS}{${S}^5$\xspace}

\newcommand{\ratiogc}{\texttt{ratio\_gc}\xspace}

\newcommand{\nspecies}{22\xspace}
\newcommand{\nafe}{\ensuremath{[\mathrm{Na}/\mathrm{Fe}]}\xspace}
\newcommand{\cplusn}{\ensuremath{[(\mathrm{C+N})/\mathrm{Fe}]}\xspace}
\newcommand{\alphafe}{\ensuremath{[\alpha/\mathrm{Fe}]}\xspace}

\usepackage{natbib}
\usepackage{hyperref}                     

\begin{document}

\title{Chemical abundances of
a Galactic Centre ejecta discovered in DESI}


   \author{Manuel Cavieres\inst{1}\fnmsep\thanks{Corresponding author: Manuel Cavieres\\
          \email{cavierescarrera@strw.leidenuniv.nl}}
          \and
          Dar\'io Gonz\'alez Picos\inst{1}
          \and
          Rogelio Albarrac\'in\inst{2,3}
          \and
          Sergey E.~Koposov\inst{4,5}
          \and 
          Elena Maria Rossi\inst{1}
          \and 
          Zephyr Penoyre\inst{1}
          \and
          Brian Thorsbro\inst{6,7}
          \and
          Nils Ryde\inst{6}
          \and
          Ting S. Li\inst{8,9,10}
          \and
          Christian Aganze\inst{11}
          \and
          Nasser Mohammed\inst{8,9}
          \and
          Mair\'ead E. Heiger\inst{12, 13}
          }

      \institute{Leiden Observatory, Leiden University,
              P.O. Box 9513, 2300 RA Leiden, the Netherlands
         \and
             Max-Planck-Institut f\"{u}r Astronomie, K\"{o}nigstuhl 17, D-69117 Heidelberg, Germany
         \and
             Heidelberg University, Grabengasse 1, 69117 Heidelberg, Germany
         \and
             Institute for Astronomy, University of Edinburgh, Royal Observatory, Blackford Hill, Edinburgh EH9 3HJ, UK
         \and
             Institute of Astronomy, University of Cambridge, Madingley Road, Cambridge CB3 0HA, UK
         \and
            Division of Astrophysics, Department of Physics, Lund University, Sweden
         \and
            Université Côte d'Azur, Observatoire de la Côte d'Azur, Laboratoire Lagrange, France.
         \and
            Department of Astronomy and Astrophysics, University of Toronto, 50 St. George Street, Toronto ON, M5S 3H4, Canada
         \and
            Dunlap Institute for Astronomy \& Astrophysics, University of Toronto, 50 St George Street, Toronto, ON M5S 3H4, Canada
         \and
            Data Sciences Institute, University of Toronto, 17th Floor, Ontario Power Building, 700 University Ave, Toronto, ON M5G 1Z5, Canada
         \and
            Kavli Institute for Particle Astrophysics \& Cosmology, Stanford University, Stanford, CA 94305, USA
         \and
            Department of Astronomy, University of Virginia, 530 McCormick Road, Charlottesville, VA 22904, USA
         \and
            The NSF-Simons AI Institute for Cosmic Origins, USA.
             }
   \date{Received January 23, 2026; accepted March 7, 2026}

 
  \abstract
  {
  Chemical abundance measurements in the Galactic Centre (GC) are extremely challenging because of high extinction, crowding, and distance. They are therefore typically limited to the brightest AGB and M giant stars, observed in the near-infrared. In contrast, hypervelocity stars (HVSs) ejected from the GC by \sag carry their natal composition into accessible regions such as the Milky Way halo, making them unique probes of GC chemistry. Here we present high-resolution optical Magellan/MIKE spectroscopy of \candidate, a bound HVS candidate in the halo with super-solar metallicity ($[\mathrm{Fe/H}]=0.2\pm0.1$). This sets it apart from accreted halo populations on radial orbits and is most naturally explained by having formed in the GC and undergone a Hills-mechanism ejection. Updated isochrone fits constrain its age to $7.4$--$11.3$\,Gyr, strengthening the case for a GC origin given need for the star to have formed in the galactic center to explain both the old age, high metallicity and kinematics. The Korg and Turbospectrum spectral-synthesis analyses yield $[\mathrm{X}/\mathrm{Fe}]$ ratios for \nspecies distinct elements; an independent line-by-line NLTE analysis also measures Li separately. Because \candidate is near the main sequence, its C, N, and O abundances are unaltered by dredge-up; to our knowledge, these are the first such measurements of unevolved material associated with the GC. Its abundance pattern lies where the nuclear star cluster (NSC), nuclear stellar disc (NSD), and inner-bulge sequences overlap, with a Na enhancement ($[\mathrm{Na/Fe}]=+0.16^{+0.03}_{-0.04}$) more modest than in NSC giants of similar metallicity. Chemistry alone thus cannot identify its progenitor; we nevertheless favour the NSC, given its large reservoir of old, metal-rich stars and the kinematics of our candidate. Alongside WINERED-HVS1 and the S-star S0-6/S10, whose abundances point to different progenitors, this tentatively suggests multiple progenitor populations for the S-star cluster. \candidate is a useful test case for using ejected dwarf stars to sample GC chemistry at optical wavelengths, complementary to near-infrared studies of giants.
  }


   \maketitle
\nolinenumbers

\section{Introduction}

The Galactic Centre (GC) is a unique environment, hosting the nearest supermassive black hole (SMBH) to Earth, Sagittarius $\mathrm{A^*}$ (\sag), with a mass of $4.3 \times 10^6 M_\odot$ \citep{eisenhauer2005, ghez2008, schodel2009, gillessen2009, boehle2016, gillessen2017, do2019, gravity2019, gravity2024}, surrounded by the densest star cluster in our galaxy, namely, the nuclear star cluster (NSC), extending up to 5 pc, which is embedded in a nuclear stellar disk (NSD) \citep{nsc_neumayer}. It is this combination of a highly dense environment and the presence of a SMBH that gives rise to extreme dynamics.

\begin{figure*}[htb]
    \centering
    \includegraphics[width=\linewidth]{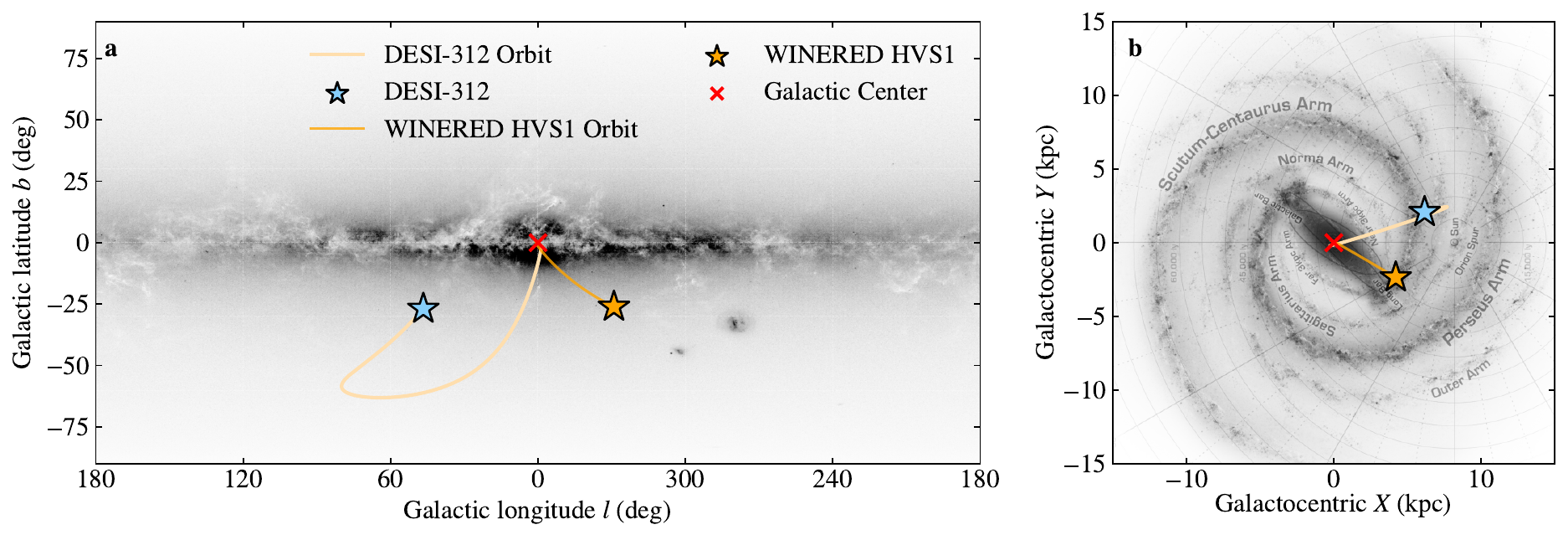}
    \caption{Sky projection in galactic coordinates (left), and bird's-eye view (right) of the backwards-integrated orbits of \candidate and WINERED-HVS1 in the Galactocentric frame, computed with mean phase-space positions. This shows the latest pericenter passage about the GC, with the stars' current position below the MW plane in accessible low-extinction regions.}
    \label{fig:orbit}
\end{figure*}

The nearest stars to \sag form a structure known as the S-star cluster, a collection of mainly B-type main-sequence stars in the innermost arcsecond of the galaxy, orbiting the SMBH on highly eccentric and inclined orbits out to 0.04\,pc \citep{ghez2008, gillessen2009, gillessen2017, gravity2018}. The origin of this cluster remains unknown because standard star formation is prohibited due to the strong tidal forces of \sag \citep{morris1993, ghez2003, genzel2010}. Given the combination of high line-of-sight extinction and source crowding direct study of the GC is challenging, and requires specialized instruments just to detect the stars in that region, such as GRAVITY \citep{gravity2017}. As a result, the chemical abundances of S-star cluster members are extremely challenging with only a single S-star measured to date \citep{Nishiyama2024S06}.

A solution to both the origin problem and the lack of chemical abundances in the S-star cluster comes from hypervelocity stars (HVS), which are stars ejected from the GC at a large velocity due to an interaction with \sag. Through a mechanism known as the Hills mechanism \citep{hills}, a progenitor binary likely originating in the NSC or NSD is tidally disrupted with a star becoming bound to the SMBH at a radii consistent with those of S-stars \citep{gouldquillen2003}, while the other star is ejected, allowing material of the GC to be observed in parts of the sky more accessible, with negligible extinction and crowding \citep{bbh_ejection, hvs_review}. 

Additionally, HVS allow us to probe not only the S-star cluster, but also the properties of their progenitor populations. Since Hills binaries can come from around the sphere of influence \citep{penoyre2025} out to several hundred pc \citep{hammer_perets_2017}, we must consider four different progenitor populations that are known to exist within a few hundred parsecs from \sag. First, the young disk surrounding the S-star cluster, often referred to as the clockwise disk (CWD), which has a top-heavy initial mass function \citep{lu2013, gallegocano2024} and is considered to have formed in a single star formation episode 2 - 6 Myr ago \citep{Paumard2006}. Secondly, the NSC is a dense cluster of predominantly old stars, with only a few percent being formed in the last billion years \citep{schodel2020}. with a canonical IMF \citep{kroupaimf} making it a typical NSC \citep{nsc_neumayer}, the combination of this population with the CWD has been proved to explain the population of HVS and S-stars \citep{verberne2025}. Outside the NSC, we have the NSD, which is a rotationally supported stellar disk, extending out to 230 pc \citep{launhardt2002, nishiyama2013, schultheis2021}, with a heterogeneous population. Finally, we must consider the inner bulge population, as the progenitor binaries for the Hills mechanism can originate from regions outside the NSD; the more general inner bulge population can also play a role in the formation of HVS.

In this paper, we present the high-resolution spectroscopic follow-up using Magellan/MIKE \citep{mike} of \candidate \citep{desihvs0}, a bound HVS found in the MW halo while having a super solar metallicity, presented in Figure \ref{fig:orbit}. As shown in \citet{desihvs0}, the chemo-dynamical properties of this object are most naturally explained by having been ejected from the GC through an interaction with \sag, and therefore, as a dynamical ejection preserves the star's natal composition, unlike the mechanisms responsible for producing the fastest runaway stars \citep[e.g.,][]{2026A&A...708A.177B}, this star should have the same chemical properties as its binary companion that was deposited in the S-star cluster. 

 A similar chemical analysis has been performed for the HVS candidate WINERED-HVS1 \citep{winered}, shown in Figure \ref{fig:orbit}. Where \citet{winered} compared its abundances with NSC, NSD and inner-bulge stars and found it broadly consistent with the NSC. However, WINERED-HVS1 is a red giant observed in the near-infrared, which limits the comparison to a handful of mainly $\alpha$-element ratios and excludes neutron-capture elements such as Eu. In contrast, \candidate is a near-main-sequence star with an optical spectrum, giving access to a wider range of elements, that are unaffected by dredge-up processes \citep[e.g.,][]{boothroyd1999, lagarde2019}, such as C, N and O.

 \candidate is particularly notable because, unlike previously identified A and B-type HVSs (such as \SSSSS HVS1, \citealt{s5hvs1}, and those reviewed by \citealt{hvs_review}), it is a G-type main-sequence star (\teff $\sim5600$K, \logg $\sim4.2$), enabling detailed chemical analysis of the atmospheres of this HVS.


Under the Hills mechanism hypothesis, the chemical abundances of these ejected stars should reflect those of their former binary companions captured into the S-star cluster. These stars therefore provide a means to constrain the origins of the S-stars by comparing their chemical abundances with those of potential progenitor populations, including the CWD, NSC, NSD, and inner bulge.


This paper is structured as follows. In Sect.~\ref{sec:methods} we describe the Magellan/MIKE observations and data reduction, the determination of the atmospheric parameters and of the elemental abundances with two independent spectral-synthesis analyses based on Korg and Turbospectrum, a line-by-line NLTE analysis, the orbit integrations, and the isochrone fit used to constrain the mass, age, and distance of \candidate. In Sect.~\ref{sec:results} we present the resulting stellar parameters and chemical abundances. In Sect.~\ref{sec:discussion} we compare the chemistry of \candidate with that of its possible progenitor populations, namely the CWD, the NSC, the NSD, and the inner bulge, as well as with the S-star S0-6/S10, and discuss the implications for the origin of the S-star cluster. Finally, we summarise our conclusions in Sect.~\ref{sec:conclusions}.

\section{Methods}\label{sec:methods}
In this section we describe the observations and methods used to determine the atmospheric parameters and the $[\mathrm{X}/\mathrm{Fe}]$ ratios reported for \nspecies distinct elements from the two joint spectral-synthesis analyses, as well as a separate line-by-line Li abundance. We also use the atmospheric parameters to improve the constraints on the star's age, mass, and distance and to update its orbit.
\subsection{Observations}\label{sec:obs}

We obtained 60 minutes of high-resolution follow-up spectroscopy of \candidate on 27 September 2025 with The Magellan Inamori Kyocera Echelle (MIKE) spectrograph \citep{mike} on the Magellan/Clay telescope with a seeing of 0.5 arcseconds. We used the 0.7 arcsecond slit; the resolving power measured from the widths of the extracted arc-lamp lines is $R\approx42{,}000$ in the blue arm and $R\approx36{,}000$ in the red arm (split at approximately 5,000\,\AA ). The data were reduced with CarPy \citep{kelson2003}, obtaining signal-to-noise ratios of about 25, 20 and 35 per pixel at 4500\,\AA, 6500\,\AA, and 7000\,\AA~ respectively with about 2 pixels per resolution element.

\subsection{Continuum, telluric correction, and order merging}\label{sec:prep}
Each echelle order was continuum normalized by dividing it by a low-pass Fourier-filtered copy of itself, which retains only variations on scales longer than ${\sim}330$ pixels. Telluric absorption in the red arm was modelled with TelFit \citep{telfit}: we mask pixels where the atmospheric transmission is below 92\%, and elsewhere divide the flux and uncertainties by the transmission, which corrects microtellurics while avoiding strongly absorbed regions. We also mask the Na\,D region (5884--5900\,\AA), which 1D LTE models do not reproduce \citep[e.g.,][]{amarsiSolarCarbonNitrogen2021}. Masking is implemented by inflating the uncertainties by a factor of $10^{6}$, so that masked pixels carry negligible weight in the likelihood. Where echelle orders overlap, we interpolate them onto the union of their wavelength grids and combine them with inverse-variance weighting,
\begin{equation}\label{eq:merge}
    f_{\lambda}
    =
    \frac{\sum_j f_{\lambda,j}\sigma_{\lambda,j}^{-2}}
         {\sum_j \sigma_{\lambda,j}^{-2}},
    \qquad
    \sigma_{\lambda}
    =
    \left(\sum_j \sigma_{\lambda,j}^{-2}\right)^{-1/2},
\end{equation}
where $j$ indexes the orders contributing at wavelength $\lambda$; pixels covered by a single order are left unchanged. These pixels are used in both the atmospheric-parameter and the abundance fits.

\subsection{Atmospheric parameter estimation}\label{sec:atm}
We fit the spectrum with 1D LTE synthetic spectra computed with Korg \citep{korg} and the Gaia-ESO line list \citep{ges_linelist}, restricted to the TSFitpy line-mask windows \citep{tsfitpy_storm, tsfitpy_gerber} for Al, Ba, Ca, Co, Cr, Eu, Fe, H, Li, Mg, Na, Ni, O, Si, Sr, Ti, and Y. Each window is padded by 5\,\AA\ on either side, and overlapping windows are merged. For a parameter vector $\boldsymbol{\theta}_{\mathrm{atm}}$, the model flux at observed wavelength $\lambda_i$ in window $w$ is
\begin{equation}\label{eq:model}
    m_i(\boldsymbol{\theta}_{\mathrm{atm}})
    =
    S_w(\lambda_i)\,
    M\!\left(\frac{\lambda_i}{1+v_r/c}\,\middle|\,\boldsymbol{\theta}_{\mathrm{atm}}\right),
\end{equation}
where $M$ is the rest-frame Korg spectrum, convolved to the resolving power of the corresponding arm (Sect.~\ref{sec:obs}), Doppler shifted by the radial velocity $v_r$ and interpolated onto the observed pixels, and $S_w$ is a cubic spline with 10 knots that absorbs residual continuum mismatch. The spline coefficients are obtained at each likelihood evaluation by non-negative weighted least squares. The log-likelihood is
\begin{equation}\label{eq:likelihood}
    \ln \mathcal{L}
    =
    -\frac{1}{2}
    \sum_{w}\sum_{i\in w}
    \left[
    \frac{
    \left(f_i - m_i(\boldsymbol{\theta}_{\rm atm})\right)^2
    }{
    b^2\sigma_i^2
    }
    +
    \ln\left(2\pi b^2\sigma_i^2\right)
    \right],
\end{equation}
where $f_i$ and $\sigma_i$ are the observed flux and uncertainty from Sect.~\ref{sec:prep}, and $b$ is a global scaling of the uncertainties that accounts for their possible underestimation and for residual modeling systematics. The parameter vector is
\begin{equation}
    \boldsymbol{\theta}_{\mathrm{atm}}
    =
    \left(
    T_{\mathrm{eff}},
    \log g,
    [\mathrm{M}/\mathrm{H}],
    [\alpha/\mathrm{H}],
    v_{\mathrm{mic}},
    v\sin i,
    v_r,
    b
    \right).
\end{equation}
Where \teff~ is the effective temperature in kelvin, \logg is surface gravity, [M/H] is the metallicity, $[\alpha/\mathrm{H}]$ is the global $\alpha$ element abundance, $v_{\mathrm{mic}}$ is the microturbulence in \kms, $v\sin i$ is the projected rotation velocity, $v_r$ is the residual radial velocity in \kms.
We adopt truncated Gaussian priors on $T_{\mathrm{eff}}$, $\log g$, $[\mathrm{M}/\mathrm{H}]$, and $[\alpha/\mathrm{H}]$, centred on 5600\,K, 4.2, 0.2, and 0.0, with widths of 250\,K, 0.3, 0.2, and 0.25\,dex, respectively; the first three are close to the DESI values for \candidate\ \citep{desimw}. For $v_{\mathrm{mic}}$ and $v\sin i$ we use truncated log-normal priors with a median of 1\kms, for $v_r$ a Gaussian prior with a width of 1\kms\ centred on an initial estimate from cross-correlation with a Korg spectrum at the DESI parameters, and for $b$ a uniform prior between 0.1 and 5. We sample the posterior with the dynamic nested sampler \texttt{dynesty} \citep{nested_sampling_1, nested_sampling_2, bounding_nested, dynesty, sergey_koposov_dynesty}; it is shown in Fig.~\ref{fig:atm}.

\subsection{Elemental abundances 1D LTE}\label{sec:korg}
We determine elemental abundances over 5000--6560\,\AA\ in a second inference pass, fixing the atmospheric parameters to \teff$=5600$\,K, \logg$=4.2$, \feh$=0.2$, $v_{\mathrm{mic}}=1$\kms, and $v\sin i=1$\kms, based on the preceding fit. The fitted parameter vector is

\begin{equation}
    \boldsymbol{\theta}_{\mathrm{abund}}
    =
    \left(
    [\mathrm{X}_1/\mathrm{Fe}],
    [\mathrm{X}_2/\mathrm{Fe}],
    \ldots,
    [\mathrm{X}_N/\mathrm{Fe}],
    v_r,
    b
    \right).
\end{equation}

where $\mathrm{X}_i/\mathrm{Fe}$ is the relative abundance of the $i$th element, with respect to the previously measured \feh. The joint fit includes 27 elemental abundance parameters (Li, C, N, O, F, Na, Mg, Al, Si, S, K, Ca, Sc, Ti, V, Cr, Mn, Co, Ni, Cu, Zn, Y, Ba, Ce, Nd, Eu, and Yb), each with a uniform prior $-0.8\leq[\mathrm{X}/\mathrm{Fe}]\leq0.8$. Six of these parameters do not yield reported Korg abundances (Sect.~\ref{sec:results}). Here one joint nested sampling run is performed over all abundance parameters, ensuring that blended regions are handled self-consistently because all species affecting the spectral region are fit together. This approach allows us to obtain robust elemental abundance measurements even in the low signal to noise regime by simultaneously fitting all spectral features present for a given species. It also allows us to fit molecular lines (see Figure \ref{fig:spectral_fits_turbo} for examples) as the abundance of each element in the molecule is also fit together, and degeneracies broken by including individual element lines.



For each trial abundance vector, the model spectrum and likelihood are computed as in Sect.~\ref{sec:atm} (Eqs.~\ref{eq:model} and \ref{eq:likelihood}), at the fixed atmospheric parameters, with the red-arm resolving power throughout, and using the order-combined pixels of Sect.~\ref{sec:prep}.

\subsection{Elemental abundances Turbospectrum}\label{sec:turbo}

A parallel, abundance analysis is performed using the Turbospectrum radiative-transfer code \citep{tsfitpy_gerber}. We adopt a fitting procedure similar to that used with Korg, fixing the fundamental stellar parameters ($T_{\rm eff}$, $\log g$, and metallicity) to their best-fit values after verifying their consistency with those inferred from Korg. We use the same Gaia-ESO line lists but fit broad spectral regions rather than narrow line-mask segments to infer molecular abundances. In particular, we derive carbon and nitrogen abundances from CH and CN molecular features. The CH molecule exhibits strong absorption around 4300\,\AA\ and is the principal optical tracer of carbon abundance in solar-type stars \citep{amarsiSolarCarbonNitrogen2021,hoppeSolarCarbonAbundance2026}. Nitrogen is constrained from CN molecular lines in the red optical spectrum \citep{amarsiSolarCarbonNitrogen2021,ryabchikovaUsingMolecularLines2022}. These CN features are intrinsically weaker and more heavily blended than those of CH, making abundance determinations from individual lines challenging. Instead, we simultaneously fit broad spectral regions (4240--4610\,\AA\ and 7710--8490\,\AA), allowing the combined imprint of numerous overlapping molecular and atomic lines to constrain the abundances while maximizing the spectroscopic information available. This spectral-synthesis approach is routinely employed for cool stars at near-infrared wavelengths, particularly M dwarfs, where the extensive blending of molecular features necessitates the simultaneous fitting of broad wavelength regions (e.g. \citealt{gonzalezpicosChemicalEvolutionImprints2025}). We infer elemental abundances by coupling the Turbospectrum forward model with a nested-sampling algorithm (PyMultiNest; \citealt{buchnerPyMultiNestPythonInterface2016}) to simultaneously fit 20 elemental abundance parameters while keeping the fundamental stellar parameters fixed. We report finite $[\mathrm{X}/\mathrm{Fe}]$ ratios for 18 of these elements in Table~\ref{tab:abundances}. The reported abundances, [C/Fe]$=-0.10\pm0.02$ and [N/Fe]$=+0.29\pm0.03$ in Figure \ref{fig:abundance_violin} and Table \ref{tab:abundances}, are inferred using this approach.

The inclusion of NLTE effects is particularly important for accurately inferring the abundances of elements such as sodium and oxygen \citep[e.g.,][]{amarsiSolarCarbonNitrogen2021}. In the joint fit, we use the NLTE coefficients for Na and O implemented in Turbospectrum \citep{tsfitpy_gerber}. The selected wavelength regions include the strong Na\,I lines around 8180\,\AA\, together with several weaker Na features, providing a more robust abundance determination than the Na\,I D doublet at 5889\,\AA. We obtain [Na/Fe]$=+0.16^{+0.03}_{-0.04}$. Similarly, the red spectral region covers the O\,I triplet at 7774\,\AA\, from which we infer [O/Fe]$=+0.04^{+0.07}_{-0.08}$.

We note that the Korg and Turbospectrum analyses employ the same line list and are therefore not fully independent.

\subsection{Elemental abundances line by line}\label{sec:linebyline}

To evaluate how NLTE affects the inferred atmospheric parameters and chemical abundances, we also include a third independent, self-consistent NLTE analysis carried out with TurboSpectrum \citep{tsfitpy_gerber, tsfitpy_storm}. This method uses an Artificial Neural Network emulator following the Payne architecture \citep{PayneYST, Kovalev2019}, trained on extensive grids of NLTE synthetic spectra generated with TurboSpectrum. The technique can provide self-consistent NLTE atmospheric parameters and abundances for 17 elements; it was introduced and validated for nearby FGKM stars in \citep{Storm2026}, then expanded to multiple types of variable stars in \citep{Albarracin2026}. It also constitutes part of the pipeline used in the 4MIDALBE-HR survey with 4MOST \citep{Bensby2019, S4paper}.

In this analysis, we adopt the methodology described in \citep{Albarracin2026}, simultaneously fitting atmospheric parameters using line masks around Mg I, Ca I, and Fe I and Fe II lines, and then determining abundances for each element on a line-by-line basis, employing the same Gaia-ESO linelist as in our earlier work. To improve the atmospheric parameter determination, we jointly fit the spectra with the photometric and astrometric constraints listed in Table \ref{tab:mistfit_photometric_measurements} using the Stellar Abundances and atmospheric Parameters Pipeline \citep[SAPP,][]{SAPP2022}. From this procedure, we infer \teff$ = 5500 \pm 50 \, \rm{K}$, \logg = $4.28 \pm 0.10$ dex, and \feh = $0.19 \pm 0.07$ dex. Using GARSTEC models \citep{GARSTEC}, SAPP yields the stellar parameters listed alongside the independent mistfit/MIST results in Table~\ref{tab:stellar_parameters_comparison}. The mass, age, metallicity, and distance estimates from the two pipelines are consistent within their quoted uncertainties.

Since this technique has been validated only over the wavelength range covered by the 4MOST HR configuration, and only for the elements computed in the original grids, we use this analysis as a complementary check on the joint spectral fits described in Sects.~\ref{sec:korg} and \ref{sec:turbo}.


\subsection{MIST models}\label{sec:mist}
We updated the mass, age, and distance estimates of \candidate\ from \citet{desihvs0} by fitting MESA Isochrones \& Stellar Tracks \citep[MIST,][]{mist1, mist2, mist3, mist4, mist5} to the spectroscopic \teff, \logg, and \feh\ (Sect.~\ref{sec:atm}), photometry from the Legacy Survey \citep{legacy_survey}, WISE \citep{wise}, 2MASS \citep{twomass}, and \gaia DR3 \citep{gaiadr3} (Table~\ref{tab:mistfit_photometric_measurements}), and the \gaia DR3 parallax. We used MIST v1.2 tracks with $v/v_{\rm crit}=0.4$ and $[\alpha/\mathrm{Fe}]=0$, interpolated using \software{minimint} \citep{minimint}.

We fitted five parameters (initial mass, log-age, metallicity, distance, and reddening) using \software{mistfit}\footnote{https://github.com/mncavieres/mistfit} \citep{desihvs0} and the dynamic nested sampler \software{dynesty} \citep{dynesty}. The Gaussian likelihood includes the photometry, with uncertainties increased by 0.1\,mag to account for systematics, spectroscopic \teff\ and \logg, and the zero-point-corrected \gaia parallax ($0.29\pm0.08$\,mas; \citealt{lindegren2021}). We adopted a Salpeter mass prior \citep{salpeter1955} over $0.1$--$100\,\mathrm{M}_\odot$, a uniform log-age prior over $\log_{10}(\mathrm{age}/\mathrm{yr})=8.5$--$10.15$, a Gaussian metallicity prior centred on the spectroscopic value and truncated to the MIST grid, and uniform priors on distance and reddening, with $0\leq E(B-V)\leq E(B-V)_{\rm SFD}+0.5$\,mag.

Relative to \citet{desihvs0}, we include the \gaia\ parallax as a Gaussian term in the likelihood instead of converting it into a prior on distance; given the 28\% fractional parallax uncertainty, a Gaussian distance prior would misrepresent the measurement.
 
 Three independent runs (different random seeds, 5000 live points each) agreed within 0.07\,$\sigma$, and we report the one with the highest evidence. The resulting unimodal mistfit/MIST posterior is summarised in Table~\ref{tab:stellar_parameters_comparison}, together with the SAPP/GARSTEC results, and shown in Fig.~\ref{fig:posterior_mist}.

Finally, we tested the sensitivity to the age prior using two alternatives: a uniform equivalent evolutionary phase (EEP) prior, as in \textsc{MINESweeper} \citep{cargile2020}, and a uniform linear-age prior, corresponding to a constant star formation rate. Both shifted all five parameters by at most 0.18\,$\sigma$ (relative to the half-width of the adopted 68\% intervals), while increasing the 16th-percentile age from 7.4 to 7.8 and 8.1\,Gyr, respectively. Thus, our adopted uniform log-age prior provides the most conservative lower age bound for \candidate.

\begin{table}[ht]
  \centering
  \caption{Photometric measurements used in the MIST spectro-photometric fit. $E(B-V)_{\rm SFD}$ is from the \citet{sfd_og} dust map with the \citet{sfd_recal} recalibration.}
  \label{tab:mistfit_photometric_measurements}
  \begin{tabular}{lc}
    \hline
    Band & Magnitude \\
    \hline
    \gaia G        & $17.125 \pm 0.003$ \\
    \gaia BP       & $17.573 \pm 0.011$ \\
    \gaia RP       & $16.531 \pm 0.006$ \\
    DECam $g$      & $17.662 \pm 0.001$ \\
    DECam $r$      & $17.010 \pm 0.001$ \\
    DECam $i$      & $16.842 \pm 0.001$ \\
    DECam $z$      & $16.772 \pm 0.001$ \\
    WISE W1        & $15.31 \pm 0.04$ \\
    WISE W2        & $15.38 \pm 0.13$ \\
    2MASS $J$      & $15.95 \pm 0.11$ \\
    2MASS $H$      & $15.37 \pm 0.13$ \\
    2MASS $K_s$    & $15.40 \pm 0.20$ \\
    $E(B-V)_{\rm SFD}$ & $0.08590$ \\
    \hline
  \end{tabular}
\end{table}

\begin{table}[!htbp]
\centering
\caption{Stellar parameters of \candidate\ from the independent spectro-photometric SAPP/GARSTEC and mistfit/MIST analyses. The column headings identify the fitting pipeline and stellar-evolution models. MIST values are posterior medians with 16th and 84th percentile uncertainties; SAPP values retain their reported uncertainties. An ellipsis indicates that no SAPP reddening estimate is reported.}
\label{tab:stellar_parameters_comparison}
\label{tab:GARSTEC_spectrophotometric}
\label{tab:mistfit_spectrophotometric}
\small
\setlength{\tabcolsep}{3pt}
\renewcommand{\arraystretch}{1.18}
\begin{tabular}{@{}lcc@{}}
\hline\hline
Parameter & SAPP/GARSTEC & mistfit/MIST \\
\hline
Mass ($M_\odot$) & $0.98\pm0.03$ & $1.02\pm0.05$ \\
$\log_{10}(\mathrm{age}/\mathrm{yr})$ & $10.04^{+0.07}_{-0.09}$ & $9.98^{+0.07}_{-0.11}$ \\
$[\mathrm{Fe}/\mathrm{H}]$ (dex) & $0.19\pm0.07$ & $0.22^{+0.10}_{-0.09}$ \\
Distance (kpc) & $3.0\pm0.7$ & $3.2^{+0.5}_{-0.4}$ \\
$E(B-V)$ (mag) & $\ldots$ & $0.12\pm0.04$ \\
\hline
\end{tabular}
\end{table}

\subsection{Orbits}
The orbits presented in Figure \ref{fig:orbit} for \candidate and WINERED-HVS1 were obtained by integrating back their trajectories using the software \software{AGAMA \citep{agama}}. For \candidate we used the \gaia sky position and proper motion, the radial velocity from \citet{desimw}, and the median distance of our MIST posterior, $3.2$\,kpc (Sect.~\ref{sec:mist}, Table~\ref{tab:stellar_parameters_comparison}). For WINERED-HVS1 we used the \gaia sky position, proper motion and radial velocity, while the distance corresponds to the median geometric distance (\texttt{r\_med\_geo}) from \citet{BailerJohnesDist}. 

We first transformed their sky position, distance, proper motion, and radial velocity into the Galactocentric reference frame using the Python package \software{Astropy \citep{astropy:2013, astropy:2018, astropy:2022}}. This transformation assumes a distance to Sgr~A* of 8.122~\kpc \citep{gravity2018}, a solar three-dimensional velocity of (12.9, 245.6, 7.78)~km\,s$^{-1}$ \citep{reid2004, drimmel2018, gravity2018}, and a solar height above the Galactic plane of 20.8~pc \citep{bennett2019}.

We adopted the \texttt{MilkyWayPotential2022} model from \software{gala} \citep{gala}, a mass model composed of a spherical nucleus and bulge, three Miyamoto--Nagai disc components \citep{MN_disk} that approximate an exponential disc, and a Navarro--Frenk--White dark matter halo \citep{NFW}. The model is calibrated to reproduce both the Milky Way rotation curve measured by \citet{2019ApJ...871..120E} and the vertical structure inferred from the solar-neighbourhood phase spiral by \citet{2023ApJ...955...74D}. 

We integrate 100,000 steps over 87\,Myr, corresponding to the latest pericenter passage, for \candidate and 15\,Myr for WINERED-HVS1. These nominal orbits illustrate the trajectories; we do not propagate observational uncertainties.


\section{Results}\label{sec:results}
In this section we present the results from the high-resolution follow-up to \candidate. 

The MIKE spectrum gives a barycentric radial velocity of $-171.75\pm0.02\,\mathrm{km\,s^{-1}}$ (Sect.~\ref{sec:atm}), differing from DESI by only $0.03\,\mathrm{km\,s^{-1}}$. We therefore find no radial-velocity evidence for binarity between these observations that would compromise the orbit used to select \candidate as an HVS \citep{desihvs0}.

Our MIKE/Korg atmospheric fit yields $T_{\mathrm{eff}}=5600\pm50$\,K, $\log g=4.20\pm0.10$, and $[\mathrm{M}/\mathrm{H}]=+0.2\pm0.1$. For comparison, DESI reports $T_{\mathrm{eff}}=5579\pm29$\,K, $\log g=4.21\pm0.07$, and $[\alpha/\mathrm{Fe}]=+0.056\pm0.022$ for \candidate\ \citep{desimw}. For comparison, the recalibrated DESI iron abundance is $[\mathrm{Fe}/\mathrm{H}]=+0.263\pm0.027$ \citep{desimw}.  In this model, Fe follows the $[\mathrm{M}/\mathrm{H}]$ scaling. Converting the fitted $[\alpha/\mathrm{H}]=0.00\pm0.22$ to the same ratio used by DESI gives $[\alpha/\mathrm{Fe}]=-0.20^{+0.24}_{-0.25}$, with the uncertainty calculated from the joint atmospheric posterior. The temperature, gravity, and iron abundance agree within the quoted uncertainties; the global $[\alpha/\mathrm{Fe}]$ estimates differ by about $1.1\sigma$.

\begin{figure}
    \centering
    \includegraphics[width=\columnwidth]{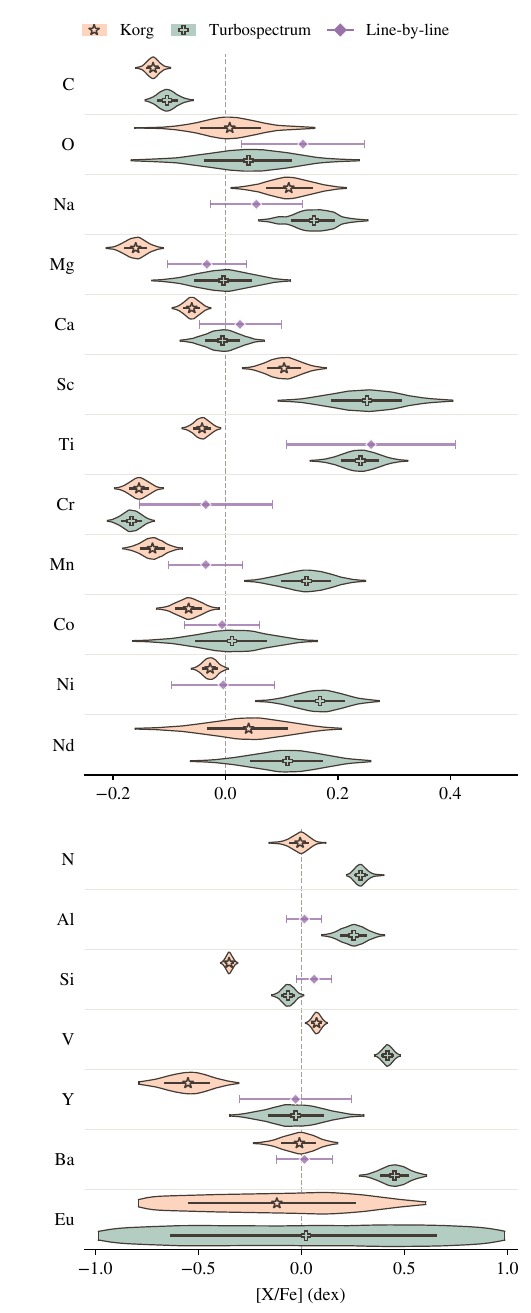}
    \caption{Korg and Turbospectrum posteriors of $[\mathrm{X}/\mathrm{Fe}]$ for \candidate, with the line-by-line measurements and their uncertainties. Violins show the central 99\% of each posterior, scaled to its peak density; bars and symbols mark the 16th--84th percentile interval and the median.}
    \label{fig:abundance_violin}
\end{figure}

Figure~\ref{fig:abundance_violin} compares the posterior distributions of elemental abundances obtained from our two joint spectral-synthesis analyses, performed with Korg and Turbospectrum, as described in Sections~\ref{sec:korg} and \ref{sec:turbo}, together with the line-by-line measurements (Sect.~\ref{sec:linebyline}). Defining the combined uncertainty as the quadrature sum of the half-widths of the respective 68\% credible intervals, we find that the two analyses agree within $1\sigma$ for O, Na, Cr, and Nd, and within $2\sigma$ for C, Ca, and Co. Mg and Sc exhibit discrepancies of $2$--$3\sigma$ ($0.15$--$0.16$\,dex), while N, Si, Ti, V, Mn, Ni, Y, and Ba show differences exceeding $3\sigma$ ($0.20$--$0.52$\,dex). The line-by-line measurements agree with the joint Turbospectrum fit within $2\sigma$ for most elements; Al, Mn, and Ba differ by approximately $2.1$, $2.1$, and $2.8\sigma$, respectively, see Table~\ref{tab:abundances}. For C, we retain the molecular-band determination (Sect.~\ref{sec:turbo}).

These discrepancies likely arise from a combination of factors. Mainly, the fact that Korg and joint Turbospectrum analyses cover different wavelength ranges, sampling distinct spectral lines with varying strengths and signal-to-noise ratios. Furthermore, although Korg and Turbospectrum produce broadly consistent synthetic spectra, systematic differences between their predictions have been documented (see \citet{korg} for a comparison). These differences highlight the importance of performing independent analyses to assess the robustness of the inferred abundances and identify potential systematic uncertainties. Consequently, for elements exhibiting discrepancies greater than $1\sigma$, we retain both abundance determinations when interpreting the chemical composition of our target, while for the case of Ti, we favour the Turbospectrum result since it also aligns with the line-by-line measurement described in section \ref{sec:linebyline}. Both measurements are therefore included in Figures~\ref{fig:na_ti_alpha}, \ref{fig:comp1}, and \ref{fig:comp2}, allowing our conclusions to account for the differences between the two analyses.

As discussed in Sect.~\ref{sec:turbo}, we utilize the Turbospectrum-based C, N, O, and Na measurements as they are more reliable given the NLTE treatment of Na and O and the use of molecular transitions for C and N. Elements for which any pair of analyses differs by more than $3\sigma$ should be regarded with caution in the following sections. Of the 27 elements in the Korg fit (Sect.~\ref{sec:korg}), the 68\% posterior intervals of Li, F, K, Yb, and Eu are at least 70\% as wide as that of the prior, whereas for every other element the interval is less than half as wide. The Al posterior instead reaches the upper prior bound. We therefore report Korg abundance ratios for 21 elements and mark these six parameters as unconstrained in Table~\ref{tab:abundances}. Turbospectrum provides finite ratios for 18 elements, including Al and 17 of the reported Korg elements. The two joint analyses thus provide $[\mathrm{X}/\mathrm{Fe}]$ ratios for \nspecies distinct elements. The line-by-line analysis provides 14 measurements: 13 $[\mathrm{X}/\mathrm{Fe}]$ ratios, including 12 of the reported Korg elements and Al, plus Li on the $A(\mathrm{Li})$ scale. The posterior medians and 68\% intervals from both joint analyses are listed in Table~\ref{tab:abundances}, together with the line-by-line measurements and their quoted uncertainties, which are also shown in Fig.~\ref{fig:abundance_violin}.

\begin{table}[!htbp]
  \centering
  \caption{Elemental abundances $[\mathrm{X}/\mathrm{Fe}]$ of \candidate\ from the Korg
  (Sect.~\ref{sec:korg}), Turbospectrum (Sect.~\ref{sec:turbo}), and line-by-line analyses.
  Korg and Turbospectrum values are posterior medians with uncertainties from the 16th and 84th percentiles; the Korg
  percentiles use the nested-sampling posterior weights. Species whose 68\% interval is at
  least 70\% as wide as that of the prior, or that reach a prior bound, are listed as unconstrained, and an ellipsis marks species
  without a reported finite measurement in the corresponding analysis. The line-by-line
  column gives the fitted central values and their quoted uncertainties. Lithium in this
  column is reported as $A(\mathrm{Li})=\log_{10}(n_{\mathrm{Li}}/n_{\mathrm{H}})+12$,
  rather than $[\mathrm{Li}/\mathrm{Fe}]$.}
  \label{tab:abundances}
  \small
  \setlength{\tabcolsep}{2pt}
  \renewcommand{\arraystretch}{1.08}
  \begin{tabular}{@{}lccc@{}}
    \hline\hline
    Species & Korg & Turbospectrum & Line-by-line \\
    \hline
    Li & unconstrained & \ldots & $1.48\pm0.15$ \\
    C\tablefootmark{b} & $-0.13\pm0.01$ & $-0.10\pm0.02$ & \ldots \\
    N\tablefootmark{b} & $-0.01^{+0.04}_{-0.05}$ & $+0.29\pm0.03$ & \ldots \\
    O\tablefootmark{b} & $+0.01\pm0.05$ & $+0.04^{+0.07}_{-0.08}$ & $+0.14\pm0.11$ \\
    F & unconstrained & \ldots & \ldots \\
    Na\tablefootmark{b} & $+0.11\pm0.04$ & $+0.16^{+0.03}_{-0.04}$ & $+0.06\pm0.08$ \\
    Mg & $-0.16\pm0.02$ & $0.00\pm0.05$ & $-0.03\pm0.07$ \\
    Al & unconstrained & $+0.25\pm0.06$ & $+0.02\pm0.09$ \\
    Si & $-0.35\pm0.02$ & $-0.06\pm0.03$ & $+0.06\pm0.09$ \\
    S & $-0.22^{+0.18}_{-0.25}$ & \ldots & \ldots \\
    K & unconstrained & \ldots & \ldots \\
    Ca & $-0.06\pm0.01$ & $-0.01\pm0.03$ & $+0.03\pm0.07$ \\
    Sc & $+0.10\pm0.03$ & $+0.25\pm0.06$ & \ldots \\
    Ti & $-0.04\pm0.01$ & $+0.24\pm0.03$ & $+0.26\pm0.15$ \\
    V & $+0.07\pm0.02$ & $+0.42\pm0.02$ & \ldots \\
    Cr & $-0.15\pm0.02$ & $-0.17\pm0.02$ & $-0.03\pm0.12$ \\
    Mn & $-0.13\pm0.02$ & $+0.14\pm0.04$ & $-0.03\pm0.07$ \\
    Co & $-0.07\pm0.02$ & $+0.01\pm0.06$ & $-0.01\pm0.07$ \\
    Ni & $-0.03\pm0.01$ & $+0.17\pm0.04$ & $0.00\pm0.09$ \\
    Cu & $-0.13\pm0.04$ & \ldots & \ldots \\
    Zn & $-0.04^{+0.24}_{-0.26}$ & \ldots & \ldots \\
    Y & $-0.55^{+0.10}_{-0.11}$ & $-0.03\pm0.13$ & $-0.03\pm0.27$ \\
    Ba & $-0.01\pm0.08$ & $+0.45^{+0.06}_{-0.07}$ & $+0.02\pm0.14$ \\
    Ce & $-0.49^{+0.22}_{-0.20}$ & \ldots & \ldots \\
    Nd & $+0.04\pm0.07$ & $+0.11\pm0.06$ & \ldots \\
    Eu & unconstrained & unconstrained & \ldots \\
    Yb & unconstrained & \ldots & \ldots \\
    \hline
  \end{tabular}
  \tablefoot{
    \tablefoottext{b}{The Turbospectrum value is adopted in this work (Sect.~\ref{sec:turbo});
    Na and O are from the NLTE synthesis.}
  }
\end{table}

The adopted Turbospectrum abundances are $[\mathrm{C}/\mathrm{Fe}]=-0.10\pm0.02$, $[\mathrm{N}/\mathrm{Fe}]=+0.29\pm0.03$, and $[\mathrm{O}/\mathrm{Fe}]=+0.04^{+0.07}_{-0.08}$. Combining C and N in linear number-abundance space, rather than averaging their logarithmic ratios, and adopting $\log\epsilon_\odot(\mathrm{C})=8.43$ and $\log\epsilon_\odot(\mathrm{N})=7.83$ from \citet{Asplund2009}, gives \cplusn{} $=+0.01\pm0.02$, shown in Fig.~\ref{fig:comp1}. We use this combined ratio when comparing with evolved Galactic-centre samples because C+N is approximately conserved through first dredge-up, even though C and N separately are not. Because \candidate\ is still near the main sequence, its surface C, N, and O abundances have not been altered by first dredge-up \citep{boothroyd1999,lagarde2019}; under the Galactic-centre-ejection interpretation, these therefore constitute, to our knowledge, the first simultaneous C, N, and O measurements of unevolved material associated with the Galactic Centre.

Compared with \citet{desihvs0}, the MIST fit gives a slightly lower mass and an older median age. The 68\% age interval changes from approximately $5.1$--$11.0$ to $7.4$--$11.3$\,Gyr, mainly raising the lower bound; the alternative priors in Sect.~\ref{sec:mist} give higher lower bounds still. The revised distance is $3.2^{+0.5}_{-0.4}$\,kpc (Table~\ref{tab:stellar_parameters_comparison}), and the 68\% intervals overlap with the previous estimates for all five fitted parameters. Changes in the spectroscopic constraints, photometry, and parallax treatment prevent us from attributing the shifts to the MIKE spectrum alone. The independent SAPP/GARSTEC estimates of mass, age, metallicity, and distance agree with the MIST results within the quoted uncertainties (Table~\ref{tab:stellar_parameters_comparison}). The old age and super-solar metallicity strengthen the case for an origin in the inner Galaxy.

Adopting the updated MIST distance of $3.2^{+0.5}_{-0.4}$\,kpc instead of the \textit{Specdis} value of $4.78\pm0.83$\,kpc used in \citet{desihvs0} substantially tightens the Galactic-centre origin of \candidate. Repeating the Monte Carlo orbit integration of \citet{desihvs0} in the same potential, the fraction of realisations whose first disc crossing lies within 1\,kpc of the GC rises from $\ratiogc=0.62$ to $0.99$. The lower distance also lowers the Galactocentric velocity, so the star reaches the GC sooner: the flight time to the closest approach to the GC is reduced from $211\pm93$\,Myr to $87^{+19}_{-12}$\,Myr. Assuming an ejection at 3\,pc from \sag, the median ejection velocity decreases from $698^{+35}_{-27}$ to $654^{+10}_{-6}$\,km\,s$^{-1}$, which remains well within the regime of Hills-mechanism ejections. 

\section{Discussion}\label{sec:discussion}
In this section we compare the chemical abundances of \candidate with those of the most likely progenitor populations and draw implications for the S-star cluster population. Working under the hypothesis that this is a hypervelocity star ejected from the galactic center through the Hills mechanism, supported by the evidence shown in \citep{desihvs0}.

\begin{figure}
    \centering
    \includegraphics[width=\linewidth]{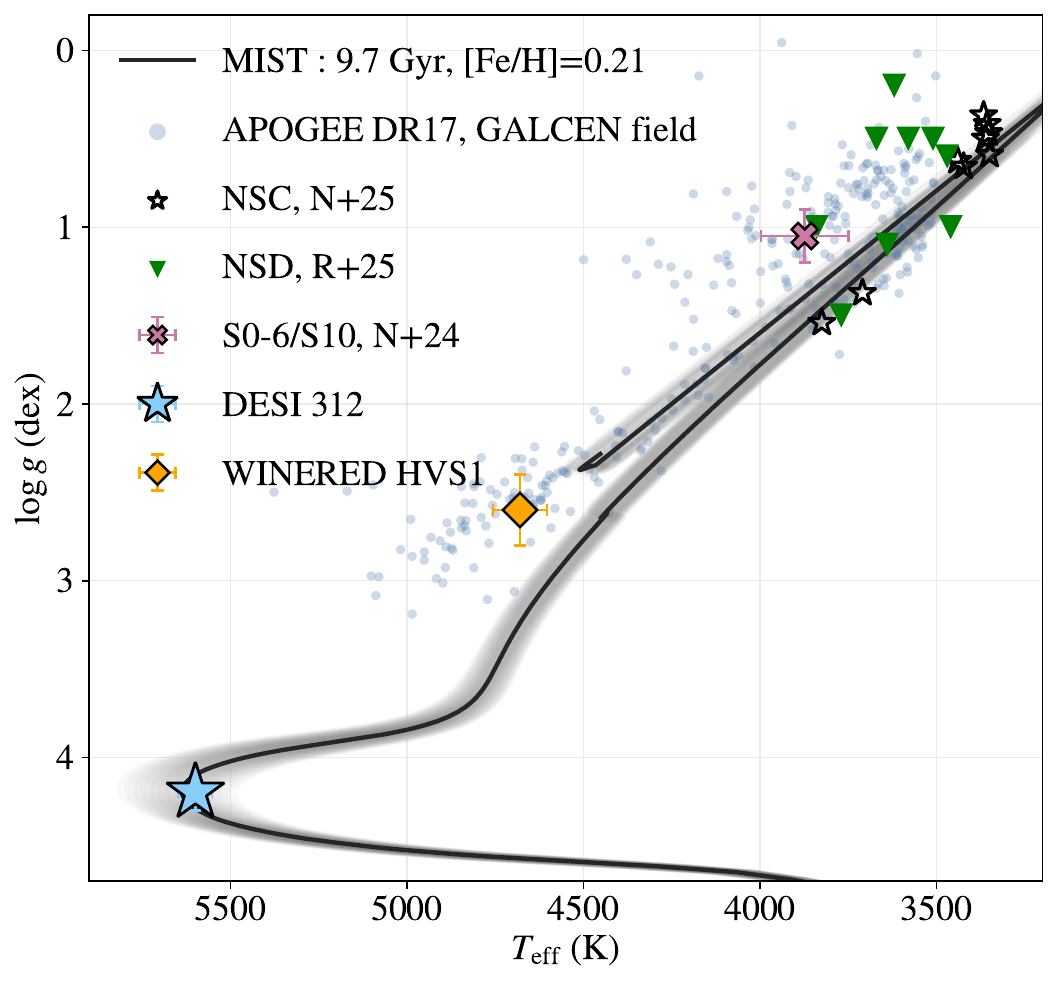}
    \caption{Kiel diagram of \candidate (blue star) and the comparison samples: APOGEE DR17 GALCEN-field giants (blue circles; \citealt{apogee_dr17}), NSC and NSD giants (black stars and green inverted triangles; \citealt{nsc_chem,nsd_chem}), WINERED-HVS1 (orange diamond; \citealt{winered}), and S0-6/S10 (magenta cross; \citealt{Nishiyama2024S06}), whose error bars span its two order-specific solutions. The black curve and grey shading show the median MIST isochrone of \candidate and its 16th--84th percentile range.}
    \label{fig:kiel}
\end{figure}

We warn the reader that in order to do a comparison we are relying on measurements from \citet{apogee_dr17, Nishiyama2024S06, nsc_chem, nsd_chem, nsc_xshooter, winered},  performed in the near infrared on red giant branch stars (see Figure \ref{fig:kiel}) using a variety of line-lists, which could induce unknown systematics. This is a limitation given by the fact that \candidate is the first main-sequence star with measured chemistry associated with the inner populations of the GC.

\subsection{The Clockwise Disk}
A combination of stars originating from the CWD and the NSC has been proposed to explain the observed populations of S-stars and HVSs \citep{verberne2025}, reflecting the need for progenitors spanning both young and old stellar populations. The CWD therefore represents a plausible progenitor population for young HVSs in general. However, a CWD origin can be ruled out for both \candidate\ and WINERED-HVS1 on the basis of their ages alone. The CWD is associated with a recent episode of star formation $\sim2$--$6$ Myr ago \citep{Paumard2006}, whereas \candidate\ has a lower age limit of 7.4 Gyr and WINERED-HVS1 is expected to be $\sim10$ Gyr old.

\subsection{Nuclear Star Cluster}

The NSC has a half-light radius of $4.2\pm0.4$\,pc and a mass of $(2.5\pm0.4)\times10^7\,\mathrm{M}_\odot$ \citep{schodel2014, nsc_neumayer}, and is considered a likely reservoir of progenitor binaries for the Hills mechanism, particularly for old stellar populations \citep{perets2007, verberne2025}. If the progenitor binary that underwent a tidal disruption with  \sag. originated in the NSC, \candidate should share the chemical properties of the NSC, additionally, the combination of the old age of our candidate (68\% interval $7.4$--$11.3$\,Gyr) and the high metallicity (\feh$=0.2\pm0.1$dex) makes the NSC the most promising progenitor. 

\begin{figure}
    \centering
    \includegraphics[width=\linewidth]{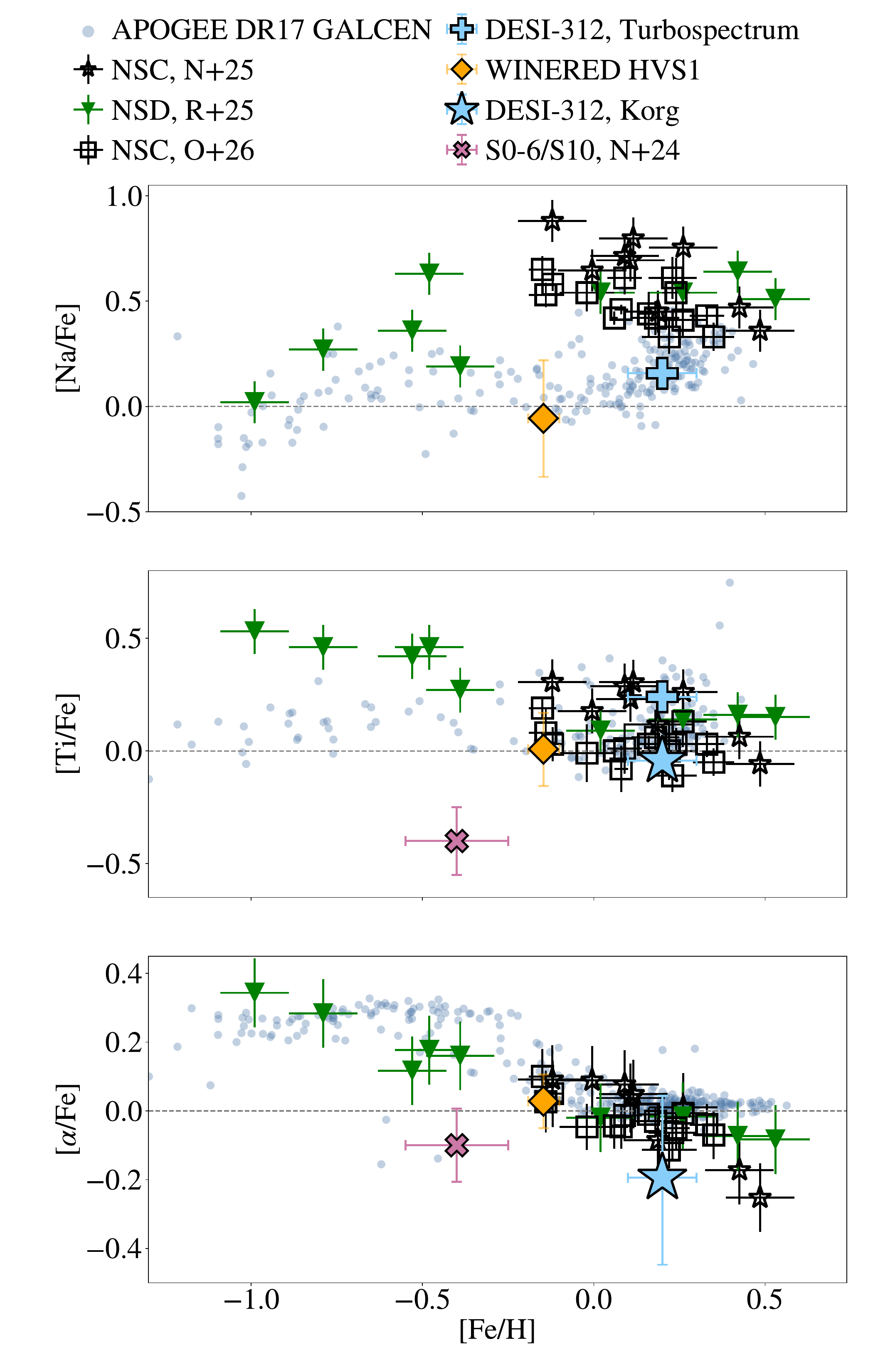}
    \caption{$[\mathrm{Na}/\mathrm{Fe}]$, $[\mathrm{Ti}/\mathrm{Fe}]$, and $[\alpha/\mathrm{Fe}]$ as a function of \feh. Symbols are as in Fig.~\ref{fig:kiel}, with black squares for the X-Shooter NSC sample \citep{nsc_xshooter}. For \candidate, stars and filled plus symbols mark the Korg and Turbospectrum values (Sect.~\ref{sec:results}), and $[\alpha/\mathrm{Fe}]$ is the global value of the Korg atmospheric fit. For the comparison stars, $[\alpha/\mathrm{Fe}]$ is the mean of $[\mathrm{Mg}/\mathrm{Fe}]$, $[\mathrm{Si}/\mathrm{Fe}]$, and $[\mathrm{Ca}/\mathrm{Fe}]$ (Mg and Ca only for S0-6/S10), or the global ASPCAP value for APOGEE. }
    \label{fig:na_ti_alpha}
\end{figure}

\begin{figure*}
    \centering
    \includegraphics[width=\linewidth]{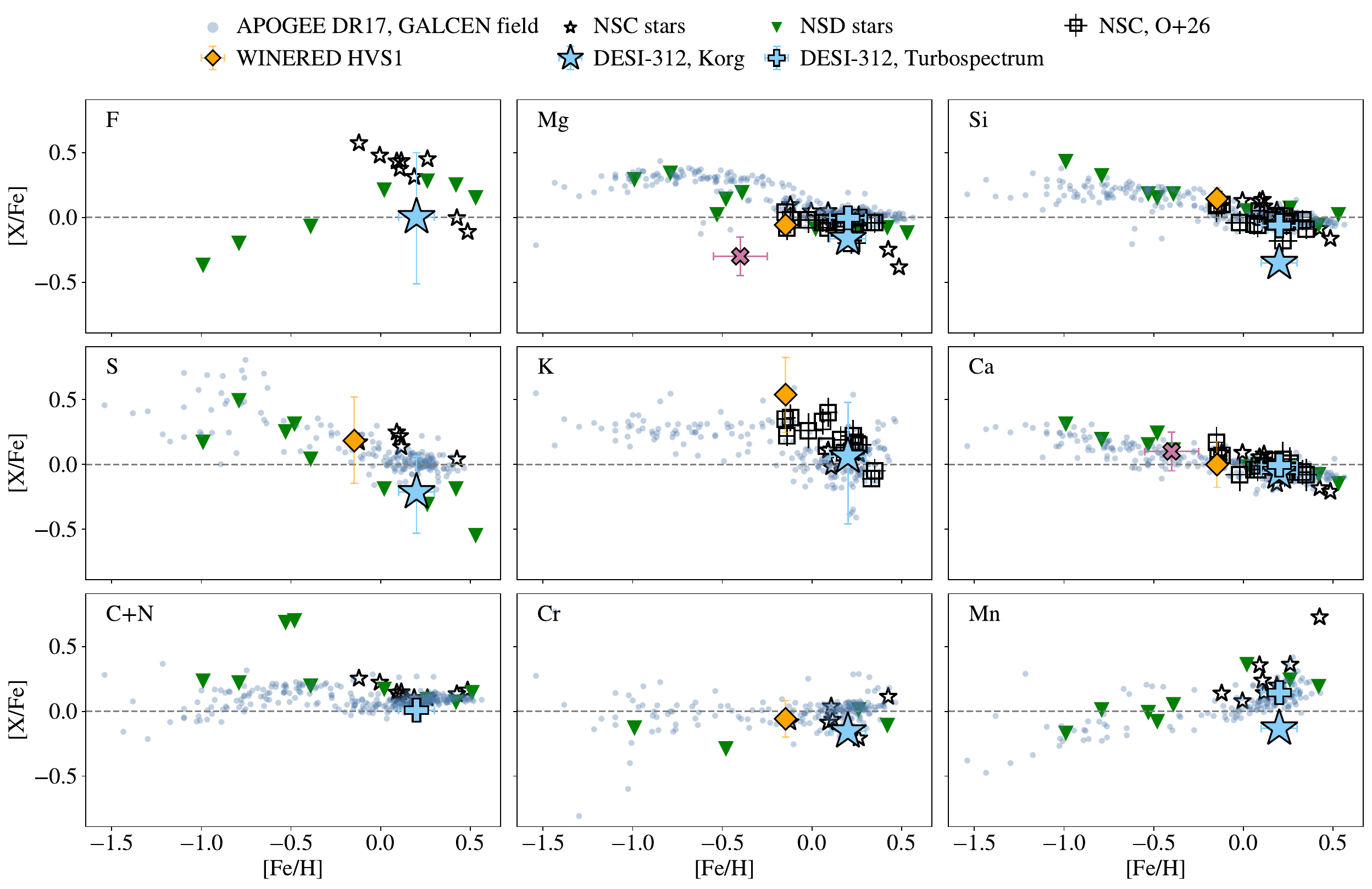}
    \caption{$[\mathrm{X}/\mathrm{Fe}]$ as a function of \feh{} for F, Mg, Si, S, K, Ca, $(\mathrm{C+N})$, Cr, and Mn. Symbols are as in Fig.~\ref{fig:na_ti_alpha}; Turbospectrum values of \candidate are shown for $(\mathrm{C+N})$ and where they differ from Korg by more than $1\sigma$. The F and K posteriors of \candidate are dominated by the prior (Sect.~\ref{sec:results}).}
    \label{fig:comp1}
\end{figure*}
\begin{figure*}
    \centering
    \includegraphics[width=\linewidth]{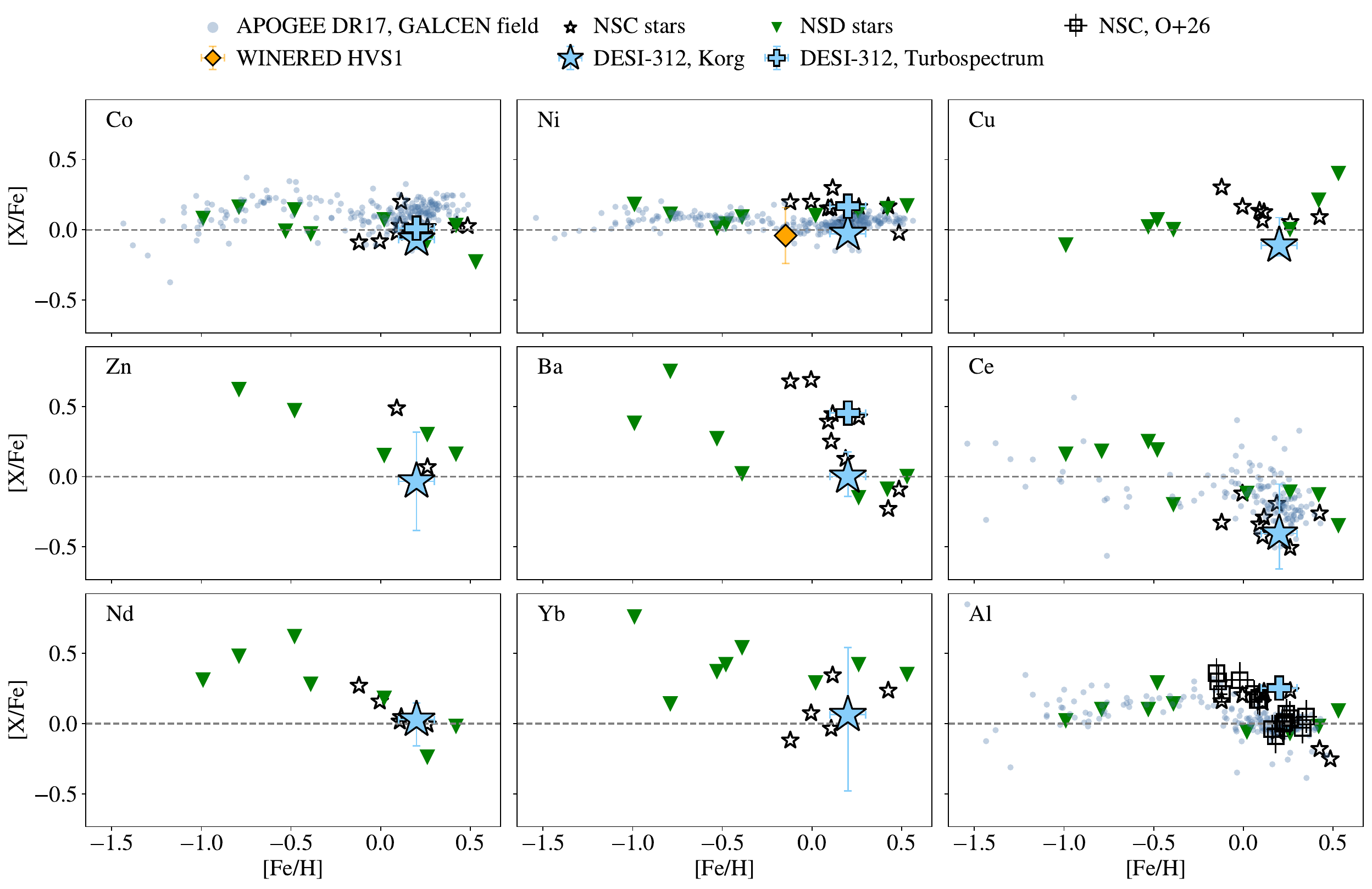}
    \caption{$[\mathrm{X}/\mathrm{Fe}]$ as a function of \feh{} for Co, Ni, Cu, Zn, Ba, Ce, Nd, Yb, and Al. Symbols are as in Fig.~\ref{fig:na_ti_alpha}; Turbospectrum values of \candidate are shown where they differ from Korg by more than $1\sigma$. The Yb posterior of \candidate is dominated by the prior (Sect.~\ref{sec:results}).}
    \label{fig:comp2}
\end{figure*}

The most distinct feature that has been found to distinguish the NSC population is an enhanced Sodium abundance at super-solar metallicity of about 0.2 dex, with most of the stars at this metallicity being around 5 - 10 Gyr in age \citep{ryde2025a, nsd_chem, nsc_chem}. Simultaneously, other abundance tracers such as the \alphafe shown in the third panel of Figure \ref{fig:na_ti_alpha} and the [Ti/Fe] trend shown in the middle panel of Figure~\ref{fig:na_ti_alpha} present no strong distinction between the NSC, NSD and inner bulge populations. 
Therefore, the \nafe sequence shown in the upper panel of Figure~\ref{fig:na_ti_alpha} is the main chemical tag for GC populations. \candidate shows a more modest Na enhancement compared to NSC giants from both samples considered, sitting at \nafe $\sim 0.16$ dex with the Korg LTE estimate giving \nafe $=+0.11$\,dex, and the additional line-by-line estimate (Sect.~\ref{sec:linebyline}) giving \nafe $=0.06 \pm 0.08$, consistent with solar, this difference of $\sim0.17$\,dex with the lowest-Na NSC stars at this metallicity still places \candidate within a sodium enhancement consistent with a NSC origin. Unlike that of WINERED-HVS1, which presents a solar sodium abundance that is $\sim0.6$\,dex below the lowest-Na NSC star at this metallicity, such a substantial difference is hard to reconcile with current NSC observations; however, there are only a handful of NSC stars. A larger sample of NSC stars will be needed to determine the origin of WINERED-HVS1, particularly given the possibility for inhomogeneity within the NSC and the lack of chemical abundances for metal poor NSC stars \citep{feldmeier2020}.

We note that NSD stars, which will be discussed in the following section, also share the sodium enhancement in the high metallicity regime (Fe/H > 0) found for NSC stars. Therefore, an Na enhancement is not a unique chemical tag for the NSC.

The Ti-enhanced NSC sequence of \citet{nsc_chem} is broadened towards lower $[\mathrm{Ti}/\mathrm{Fe}]$ by the X-Shooter sample \citep{nsc_xshooter}, increasing its overlap with the NSD and inner-bulge sequences. For \candidate, the Korg abundance, $[\mathrm{Ti}/\mathrm{Fe}]=-0.04$, disagrees with the joint Turbospectrum ($+0.24$) and line-by-line ($+0.26\pm0.15$) estimates. Although the consistency of the latter two estimates favours Ti enhancement, the discrepancy between methods prevents a firm conclusion about the Ti abundance. Moreover, the overlapping population sequences throughout the range spanned by these estimates mean that resolving this discrepancy would not, by itself, uniquely identify the progenitor population.

With current observations, DESI 312 is found to be consistent with a NSC origin, while WINERED-HVS1 is inconsistent with a NSC origin if the current sample is taken at face value, contrary to the findings reported by \citet{winered}. However, there is evidence for inhomogeneity in the chemistry on the NSC, as nuclear star clusters are not expected to be composed of a single stellar population, instead being built up from many accreted globular clusters \citep{nsc_neumayer}, or many star formation episodes with gas inflow \citep{thorsbro2025}, resulting in many stellar populations \citep[eg.,][]{omegacat9}. 

\subsection{Nuclear Stellar Disk}
The NSD is a rotationally supported stellar disc extending to $R\sim230$\,pc with a vertical scale height of $\sim45$\,pc \citep{launhardt2002, nishiyama2013, schultheis2021}, and has a total stellar mass of $\sim10^9\,\mathrm{M}_\odot$ \citep[$M_{\rm NSD}=1.05^{+0.11}_{-0.10}\times10^9\,\mathrm{M}_\odot$;][]{launhardt2002, sormani2022}. Chemically, the NSD has different chemical abundances than NSC stars. NSD stars are typically more metal-poor than NSC stars, while also spanning a wider [Fe/H] range \citep{schultheis2021, nsd_chem}; see Figs.~\ref{fig:na_ti_alpha}, \ref{fig:comp1}, and \ref{fig:comp2}. These figures also show that the NSD shares both the chemical abundance trends of the NSC at large metallicities and those of the inner bulge sequence at lower metallicities, in line with the findings of \citet{ryde2025a} that the galactic centre seems to be a continuation of the inner bulge trends discussed in the following subsection. 

\candidate has an \nafe $\sim0.4$\,dex lower than NSD stars at the same metallicity , while no NSD star has a metallicity within 0.15\,dex of WINERED-HVS1; the NSD star closest to it in \feh lies $\sim0.6$\,dex higher in \nafe. In line with the findings of \citet{winered}, we find that the chemistry of WINERED-HVS1 is inconsistent with the NSD based on the [Na/Fe] vs [Fe/H] distribution, while they reached that conclusion by comparing low [Si/Mg] of the WINERED-HVS1 with high [Si/Mg] for NSD stars.

Simultaneously, \candidate nature with respect to the NSD is less clear, as the \nafe available sample of NSD stars at super-solar metallicities is sparse; we cannot rule out a NSD origin for our candidate. Additionally, we find good alignment overall in [Ti/Fe] and $[\alpha/\mathrm{Fe}]$, the latter only within the large uncertainty of the atmospheric-fit value ($-0.20^{+0.24}_{-0.25}$), as well as all elements shown in Figures \ref{fig:comp1} and \ref{fig:comp2}, with the exception of [Si/Fe] and [Mn/Fe], which should be taken with caution given that for these elements our analyses diverge (see Figure \ref{fig:abundance_violin}). The Korg [Si/Fe] lies below all comparison samples at the metallicity of \candidate, whereas the Turbospectrum value, $[\mathrm{Si}/\mathrm{Fe}]=-0.06\pm0.03$, and the line-by-line value, $+0.06\pm0.09$, follow the NSC, NSD and inner-bulge sequences; for Mn, the Turbospectrum value makes the star consistent with the NSD trend. The Turbospectrum [Ba/Fe] ($+0.45$), also shown in Fig.~\ref{fig:comp2}, lies above the NSD stars and close to the NSC giants. However, the line-by-line value ($+0.02\pm0.14$) does not support this Ba enhancement.

\subsection{Inner Bulge}

The comparison stars shown as semi-transparent blue circles in Figures~\ref{fig:na_ti_alpha}, \ref{fig:comp1}, and \ref{fig:comp2} were drawn from APOGEE DR17 sources carrying the field designation \texttt{GALCEN} \citep{apogee_dr17}. The GALCEN pointing covers approximately $2.9\,\mathrm{deg}^{2}$, centred at $(l,b)=(0.173^\circ,-0.070^\circ)$, and combines APO and LCO observations obtained with a special target selection that included cool giants, long-period variables and AGB stars, and known or candidate supergiants \citep{schultheis2015,schultheis2020}. From 856 catalogue entries, we retained 323 unique stars after requiring the APOGEE \texttt{STAR\_BAD} bit to be unset, $S/N\geq70$, $3500\leq T_{\mathrm{eff}}\leq5500\,\mathrm{K}$, $-0.5\leq\log g\leq3.5$, and a finite, unflagged $[\mathrm{Fe} \mathrm{H}]$ measurement; for duplicate observations, the highest-$S/N$ entry was retained. Each elemental panel additionally requires a finite abundance and uncertainty and a zero element-specific abundance flag. The selected stars span approximately $-0.88^\circ<l<+0.79^\circ$ and $-0.94^\circ<b<+1.00^\circ$, corresponding, if placed at a Galactic-centre distance of $8.2\,\mathrm{kpc}$, to projected separations of approximately $5$--$156\,\mathrm{pc}$ from \sag, and therefore it is less likely to include NSC stars, but it has a high chance of including NSD stars. These projected separations do not constitute three-dimensional Galactocentric distances: the \texttt{GALCEN} designation is an angular field label, and the present selection applies no distance, extinction, proper-motion, or radial-velocity membership criterion. It is therefore broader than the distance-screened GALCEN sample of \citet{schultheis2020} and the additional photometric and kinematic selections used to identify candidate nuclear-stellar-disc stars \citep{sormani2020}. The plotted distribution should consequently be treated as a Galactic-centre line-of sight sample that can contain foreground disc, bulge/bar, and nuclear-region stars; therefore, representing the general inner-bulge region \citep{zoccali2024}. 

As previously described, the lower limit on the separation from \sag places the GALCEN sample within the NSD, however this sample does not share the enhanced Ti found for the kinematically selected NSD giants in \citet{nsd_chem}. This is expected, given that our selection samples a mixture of populations whose relative contributions have not been quantified. Hence, we will use it to assess chemical compatibility with stars observed towards the inner Galaxy. \citet{hammer_perets_2017} proves that binaries out to several hundreds pc can be placed in low angular momentum orbits required for the Hills mechanism; meaning that this inner galaxy  population gives a vast reservoir for HVS progenitor binaries. 


For both HVSs considered in this study, \candidate and WINERED-HVS1, this inner bulge sample is remarkably consistent as a progenitor, with a good alignment in all elements measured in this study for which GALCEN abundances are available, except for the Korg [Si/Fe] discussed above, and in all species measured in \citet{winered}. Most importantly, it is the only one that would explain the more modest \nafe observed for our candidate and the \nafe abundance found for WINERED-HV1. 

\subsection{Comparison with S0-6/S10}
\citet{Nishiyama2024S06} provided the first and only measurement of chemical abundances for an S-star, they obtained high resolution spectroscopy of S0-6/S10 using the Infrared Camera and Spectrograph (IRCS; \citealt{kobayashi2000}) and the AO188 adaptive optics system \citep{hayano2010} on the Subaru Telescope \citep{subaru_telescope}, they obtained measurements of [M/H], [Fe/H], [alpha/Fe], [Ca/Fe], [Mg/Fe], [Ti/Fe]), included in Figures \ref{fig:na_ti_alpha} and \ref{fig:comp1}, as well as a very old age ($\geq 10$ Gyr). As shown by the anomalous [Ti/Fe], \alphafe and [Mg/Fe] this star is unlikely to originate from the NSC, NSD or inner bulge population, only sharing a common age with the HVS included here. Hence we reject a common origin for S0-6/S10 and both HVS considered in this study, WINERED-HVS1 and \candidate. Further supporting the authors' hypothesis that this S-star has an origin different from that of stars born in the central pc region, and it is more naturally explained by being born in a dwarf galaxy and then migrated to the center of our galaxy \citep{Nishiyama2024S06}. 

\subsection{Implications on the S-star cluster}

We have compared the chemical properties of \candidate\ and WINERED-HVS1 with those of the most plausible progenitor populations in the Galactic Centre, including the CWD, NSC, NSD, and inner bulge. In the Hills mechanism, the tidal disruption of a stellar binary by \sag ejects one component as an HVS while leaving the other bound to the SMBH, potentially contributing to the S-star cluster \citep{hills, bbh_ejection, hvs_review}. Assuming that the binary components formed together and therefore share the same natal chemical composition, the abundances of an HVS can be used as a proxy for those of its captured companion. Consequently, identifying the progenitor populations of Hills-ejected HVSs also provides an indirect means of constraining the populations that contribute stars to the S-star cluster.

Our chemical comparison favours an inner-bulge origin for WINERED-HVS1. In contrast, \candidate lies in the region of abundance space where the NSD, NSC, and inner-bulge populations overlap, preventing an unambiguous chemical association. We nevertheless favour the NSC as its most likely progenitor, given the large reservoir of old, metal-rich stars \citep{nsc_chem}, theoretical predictions \citep{perets2007, Zhang2013} and the need for an old NSC component to explain the S-star and HVS populations \citep{verberne2025}. In parallel, the only S-star for which a similarly detailed chemical analysis is currently available, S0-6/S10, has been argued to originate from an accreted galaxy \citep{Nishiyama2024S06} and exhibits an abundance pattern distinct from the Galactic-centre populations considered here (Figs.~\ref{fig:na_ti_alpha} and \ref{fig:comp1}).

Taken together, \candidate, WINERED-HVS1, and S0-6/S10 provide three independent chemical probes of the populations capable of contributing stars to the S-star cluster. Although the present sample is small, their markedly different abundance patterns suggest that the S-star cluster may have a heterogeneous origin, drawing stars from multiple native Galactic-centre populations---potentially including the NSC, NSD, and inner bulge---in addition to an accreted component, as suggested by S0-6/S10. A substantially larger sample of chemically characterized S-stars and Hills-ejected HVSs will be required to determine the relative contributions of these populations.


\section{Conclusions}\label{sec:conclusions}

We have obtained and analyzed high-resolution optical Magellan/MIKE spectroscopy of the bound hypervelocity-star candidate \candidate, an old, super-solar-metallicity G-type star whose chemo-dynamical properties are consistent with an origin in the Galactic Centre \citep{desihvs0}. We analysed its optical spectrum using two independent Bayesian spectral-synthesis frameworks based on Korg and Turbospectrum, obtaining atmospheric parameters and reported $[\mathrm{X}/\mathrm{Fe}]$ ratios for 21 and 18 elements, respectively. Their combined set contains \nspecies distinct elements. An independent line-by-line NLTE analysis provided atmospheric parameters and 14 elemental measurements, including Li on the $A(\mathrm{Li})$ scale. We performed a spectrophotometric analysis based on MIST models which was also independently repeated based on SAPP/GARSTEC models, both agree and place \candidate as an old ($\gtrsim 7,\mathrm{Gyr}$) solar mass star, currently at around 3 kpc from the sun.
The combination of this old age with its confirmed super-solar metallicity further strengthens the case that \candidate originated in the inner regions of the Milky Way, as this is the region in which the constraints from the dynamics presented in \citet{desihvs0}, would allow for a metal rich star to form at an old age. 

The main-sequence evolutionary state of \candidate offers an important advantage over the giant stars typically observed in the Galactic Centre, as its surface C, N, and O abundances have not been altered by first dredge-up. Under the Galactic Centre ejection hypothesis, these measurements represent, to our knowledge, the first simultaneous C, N, and O abundance determinations in an unevolved star originating from the Galactic Centre, providing a rare constraint on the chemical composition of its interstellar medium at the time of the star's formation. However, the broad constraints on the stellar age limit our ability to establish when this chemical composition was present. Furthermore, our high-resolution optical spectrum enables the measurement of a substantially broader range of elements than previously available for HVS candidates.

We compared these abundances with the most plausible progenitor populations for Hills binaries: the CWD, NSC, NSD, and inner bulge. The old age of \candidate immediately excludes an origin in the young CWD. Chemically, however, \candidate occupies a region in which the NSC, NSD, and inner-bulge populations substantially overlap. Its modest Na enhancement and its Ti abundance are consistent with the presently observed NSC population, but neither provides a unique chemical tag. The use of \nafe as a chemical tag fails since super-solar-metallicity NSD stars can also be Na-enhanced, while regarding titanium, the inclusion of recent NSC measurements substantially broadens the observed Ti distribution to now coincide with NSD stars. We therefore cannot identify the progenitor population of \candidate from chemistry alone. Nevertheless, we consider the NSC to be its most likely origin given its large reservoir of old, metal-rich stars, theoretical expectations for the supply of Hills binaries, and the requirement for an old NSC component in models reproducing the observed HVS and S-star populations \citep{perets2007,Zhang2013,verberne2025}. An NSD or inner-bulge origin, however, cannot presently be excluded.

Applying the same comparison to WINERED-HVS1 favours an inner-bulge origin. Its approximately solar sodium abundance is difficult to reconcile with the currently measured NSC and NSD sequences at its metallicity, whereas the inner-bulge population provides a natural match to its observed abundance pattern. This conclusion remains limited by the small number of chemically characterised NSC and NSD stars, particularly at lower metallicities, and by possible abundance systematics introduced by comparing optical measurements of a dwarf with near-infrared measurements of giant stars obtained using different analysis pipelines and line lists.

These results also have implications for the origin of the S-star cluster. In the Hills mechanism, disruption of a binary by \sag ejects one component as an HVS while depositing the other onto a tightly bound orbit around the SMBH \citep{hills,bbh_ejection,hvs_review}. If the two stars share their natal chemical composition, a Hills-ejected HVS therefore provides a chemical proxy for the otherwise difficult-to-observe captured companion. Within this interpretation, \candidate is consistent with a binary supplied by the nuclear populations, most plausibly the NSC, whereas WINERED-HVS1 points towards an inner-bulge progenitor. Meanwhile, S0-6/S10, the only S-star with detailed chemical abundances currently available, exhibits a distinct abundance pattern and has been interpreted as originating in an accreted system \citep{Nishiyama2024S06}.

Taken together, these three stars provide tentative evidence that the population feeding the S-star cluster may be chemically heterogeneous, with contributions from multiple Galactic-centre populations and potentially from accreted material. The present sample is far too small to determine their relative contributions, and the overlap between the chemical sequences of the NSC, NSD, and inner bulge remains a fundamental limitation. A larger and homogeneously analysed sample of Hills-ejected HVSs, S-stars, and Galactic-centre comparison stars---together with improved neutron-capture-element abundances and secure orbital classifications---will be required to transform HVS chemical tagging into a quantitative probe of the origin and assembly history of the S-star cluster.

The forthcoming second data release of the DESI stellar survey, expected to include up to ten million stars, together with other large spectroscopic surveys such as WEAVE \citep{weave} and 4MOST \citep{4most}, will significantly expand the prospects for discovering new HVSs. Combined with the improved astrometry anticipated from the fourth data release of \gaia, these surveys are expected to substantially increase the known HVS population, enabling its statistical characterization and providing new constraints on the stellar populations and dynamical processes in the Galactic Centre.

\begin{acknowledgements}
      The authors acknowledge Bianca Sersante for their insightful discussions. RA acknowledge funding from the European Research Council (ERC) under the European Union’s Horizon 2020 research and innovation programme (Grant agreement No. 949173). EMR acknowledges support from European Research Council (ERC) grant number: 101002511/project acronym: VEGA\_P. SK acknowledges support from the Science \& Technology Facilities Council (STFC) grant ST/Y001001/1. 
      This paper includes data gathered with the 6.5-meter Magellan Telescopes located at Las Campanas Observatory, Chile. BT acknowledges support from the Kavli Institute of Particle Astrophysics and Cosmology at Stanford University and SLAC National Accelerator Laboratory.
      This work was performed using the ALICE compute resources provided by Leiden University. This work uses the The Magellan Inamori Kyocera Echelle spectrograph data gathered with the 6.5 meter Magellan Telescope located at Las Campanas Observatory, Chile.
\end{acknowledgements}

%

\bibliographystyle{bibtex/aa}
\bibliography{bib}







   
  



\begin{appendix}




\onecolumn

\section{Posterior distributions from nested sampling inference}

\begin{figure*}[ht!]
    \centering
    \includegraphics[width=\linewidth]{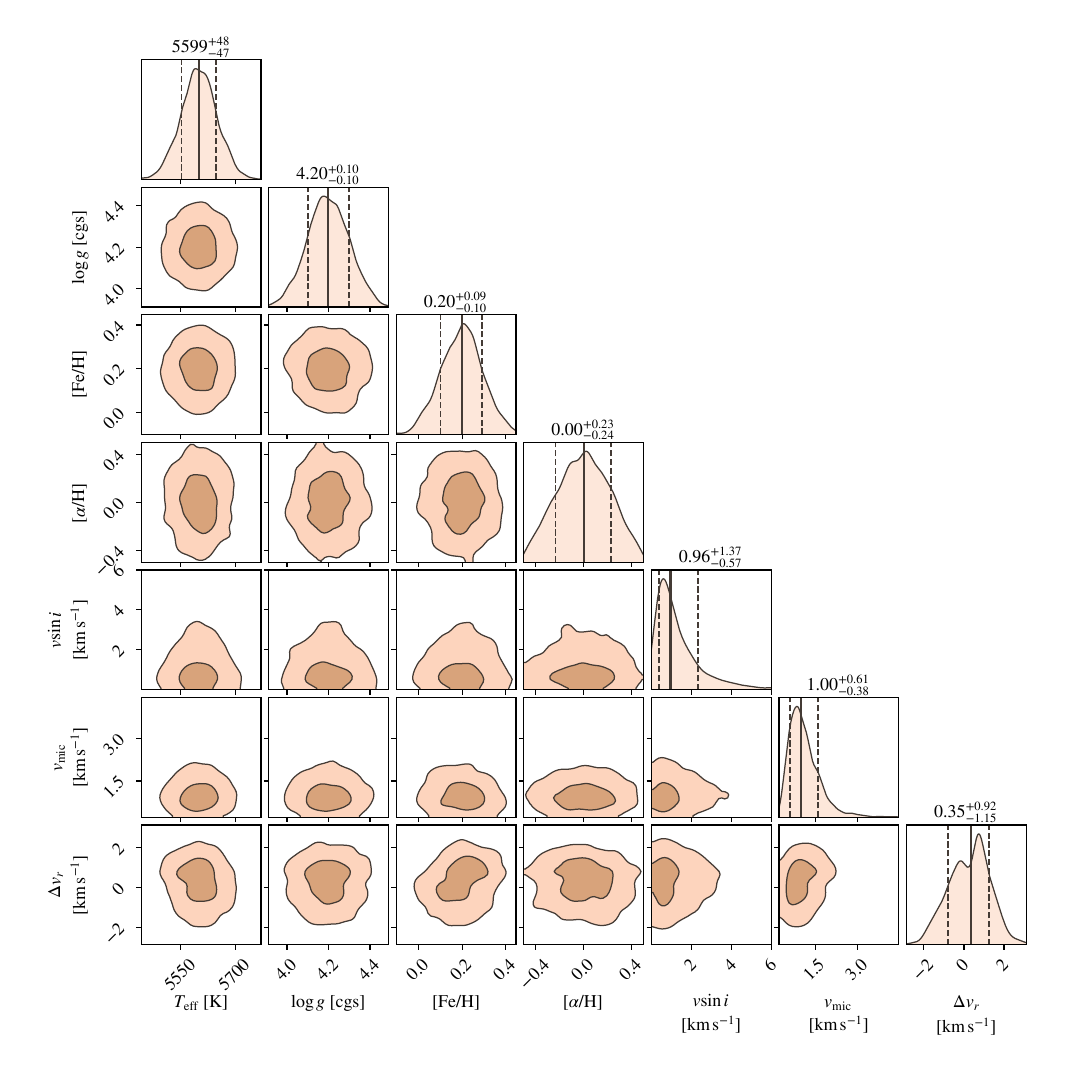}
    \caption{Posterior distribution of the atmospheric parameters for \candidate based on 1D LTE analysis with Korg on Magellan/MIKE spectra.}
    \label{fig:atm}
\end{figure*}

\begin{figure*}[ht!]
    \centering
    \includegraphics[width=\linewidth]{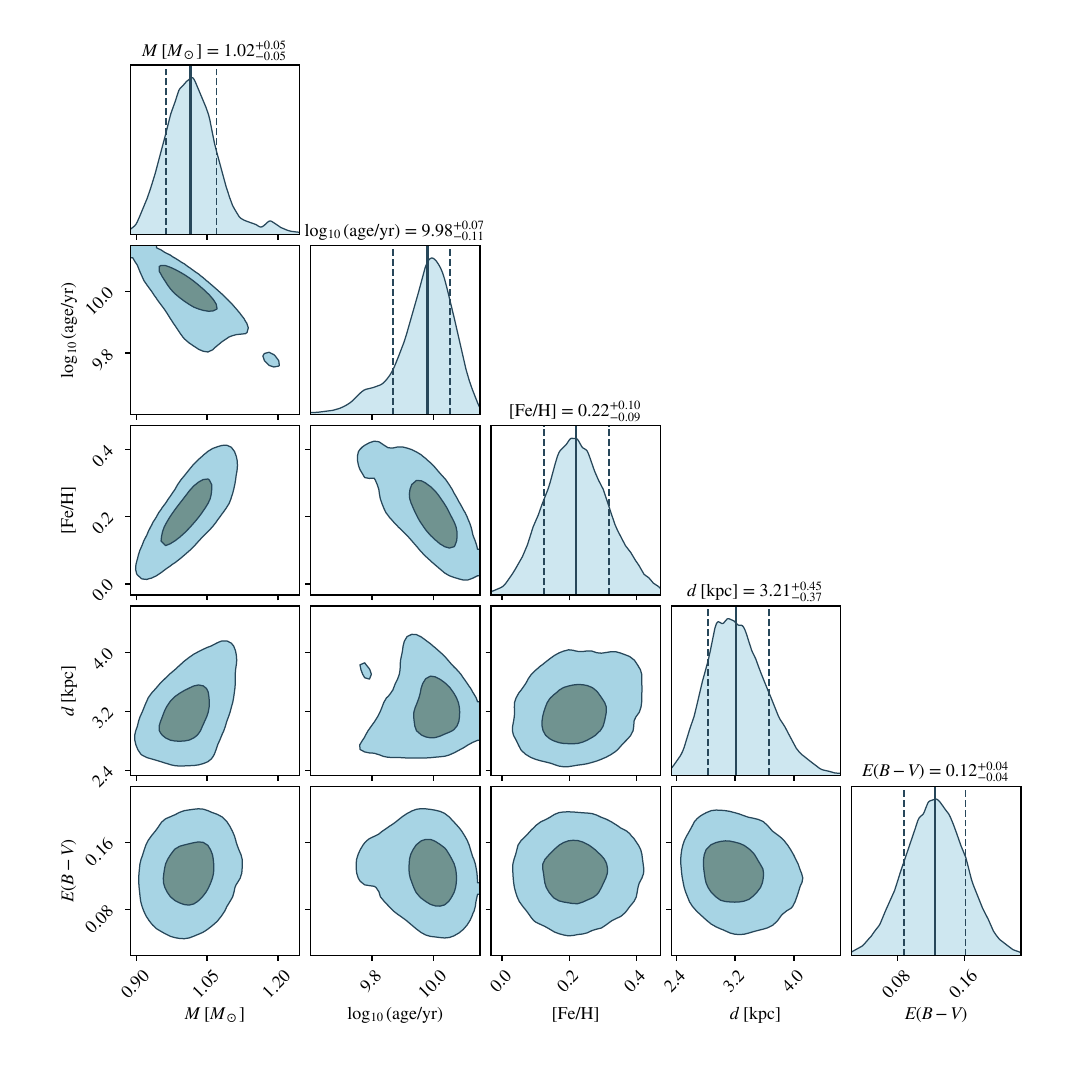}
    \caption{Posterior distribution for the spectrophotometric model using \textit{mistfit} on the photometric measurements presented in table \ref{tab:mistfit_photometric_measurements} and the atmospheric parameters obtained in section \ref{sec:atm}.}
    \label{fig:posterior_mist}
\end{figure*}

\section{Spectral fitting examples}
\begin{figure*}[ht!]
    \centering
    \includegraphics[width=\linewidth]{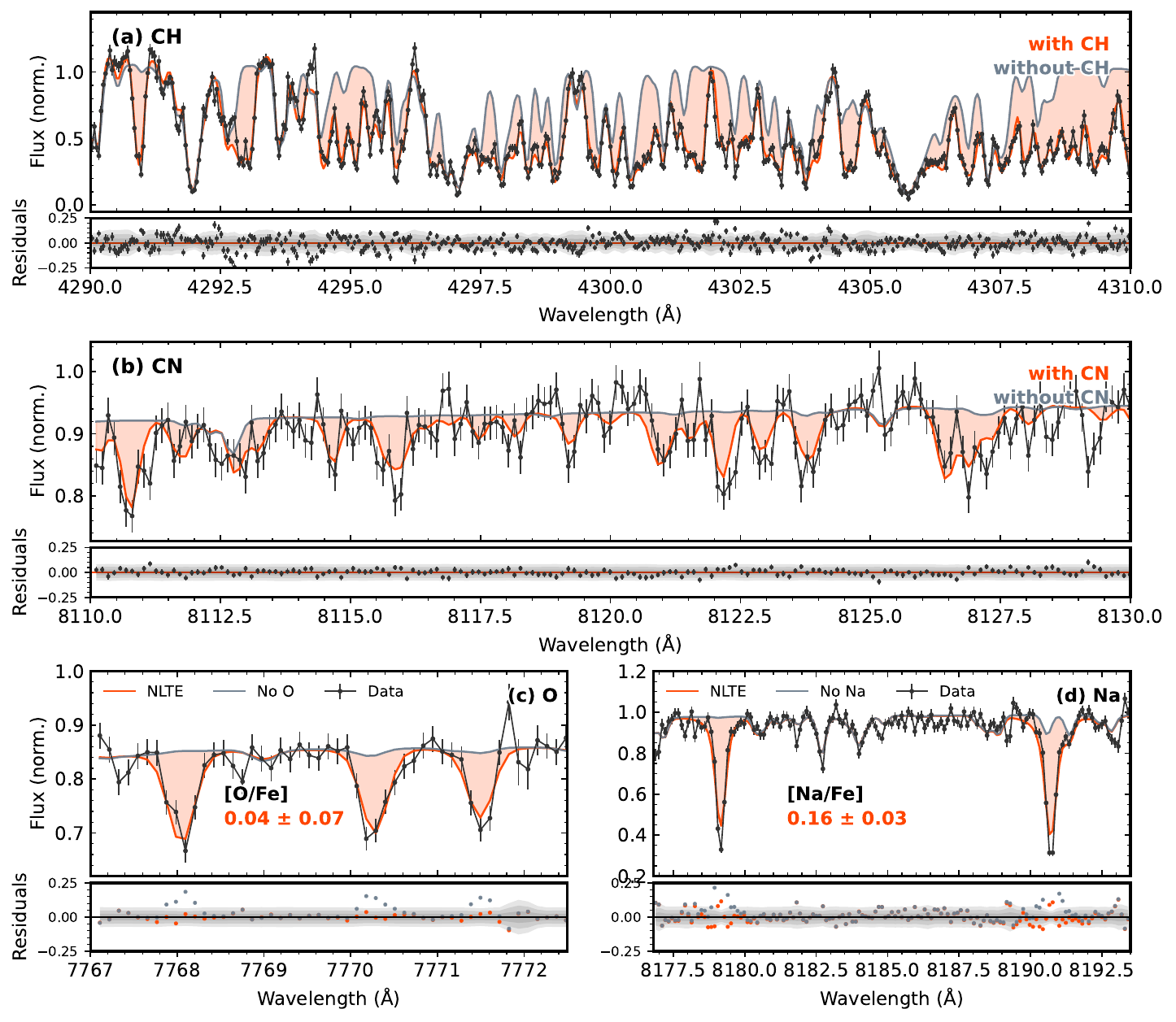}
    \caption{Comparison of the observed MIKE spectra with synthetic spectral models and their residuals. The observed spectra are shown in black, with the residuals relative to the best-fitting models displayed at the bottom of each panel. \textbf{a)} CH molecular features observed in the MIKE blue arm. The best-fitting model including CH absorption is shown in orange, while a model without CH absorption is shown in grey. The shaded orange region highlights the contribution of CH absorption. \textbf{b)} CN molecular features observed in the MIKE red arm, with models including and excluding CN absorption shown in orange and grey, respectively. \textbf{c)} The O I triplet region, showing the best-fitting NLTE models with and without the prescense of oxygen. Neglecting NLTE effects leads to an overestimation of the oxygen abundance by approximately $0.2\,\mathrm{dex}$. \textbf{d)} Strong Na absorption lines in the MIKE red arm, showing the best-fitting NLTE model alongside a model without Na absorption.  }
    \label{fig:spectral_fits_turbo}
\end{figure*}
\FloatBarrier 
\clearpage

\end{appendix}
\end{document}